# Nanoelectromechanical control of spin-photon interfaces in a hybrid quantum system on chip

Genevieve Clark,[1,2,*] Hamza Raniwala[2], Matthew Koppa,[3] Kevin Chen,[2] Andrew Leenheer,[3] Matthew Zimmermann[1], Mark Dong,[1,2] Linsen Li,[2] Y. Henry Wen[1], Daniel Dominguez,[3] Matthew Trusheim[2,4], Gerald Gilbert,[5,†] Matt Eichenfield,[6,†] Dirk Englund[2,†]

[1]*The MITRE Corporation, 202 Burlington Road, Bedford, Massachusetts 01730, USA*
[2]*Research Laboratory of Electronics, Massachusetts Institute of Technology, 50 Vassar Street, Cambridge, Massachusetts 02139, USA*
[3]*Sandia National Laboratories, P.O. Box 5800 Albuquerque, New Mexico 87185, USA*
[4]*DEVCOM, Army Research Laboratory, Adelphi, MD, 20783, USA*
[5]*The MITRE Corporation, 200 Forrestal Road, Princeton, New Jersey 08540, USA*
[6]*College of Optical Sciences, University of Arizona, Tucson, Arizona 85719*
*gclark@mitre.org  †englund@mit.edu  †eichenfield@arizona.edu  †ggilbert@mitre.org

Atom-like defects or color centers (CC's) in nanostructured diamond are a leading platform for optically linked quantum technologies, with recent advances including memory-enhanced quantum communication, multi-node quantum networks, and spin-mediated generation of photonic cluster states. Scaling to practically useful applications motivates architectures meeting the following criteria: C1 individual optical addressing of spin qubits; C2 frequency tuning of CC spin-dependent optical transitions; C3 coherent spin control in CC ground states; C4 active photon routing; C5 scalable manufacturability; and C6 low on-chip power dissipation for cryogenic operations. However, no architecture meeting C1-C6 has thus far been demonstrated. Here, we introduce a hybrid quantum system-on-chip (HQ-SoC) architecture that simultaneously achieves C1-C6. Key to this advance is the realization of piezoelectric strain control of diamond waveguide-coupled tin vacancy centers to meet C2 and C3, with ultra-low power dissipation necessary for C6. The DC response of our device allows emitter transition tuning by over 20 GHz, while the large frequency range (exceeding 2 GHz) enables low-power AC control. We show acoustic manipulation of integrated tin vacancy spins and estimate single-phonon coupling rates over 1 kHz in the resolved sideband regime. Combined with high-speed optical routing with negligible static hold power, this HQ-SoC platform opens the path to scalable single-qubit control with optically mediated entangling gates.

## I. Introduction

To reach practical utility, quantum information processors in proposed quantum repeater networks[1–7] and modular quantum computers[2,8,9] require thousands of logical qubits[9], motivating the development of quantum control architectures that can scale beyond millions of physical qubits. Solid-state atom-like defects are particularly well suited for scalability due to their potential for high-density integration in nanophotonic devices, enabling close proximity with very large-scale integrated circuit (VLSI) and photonic integrated circuit (PIC) semiconductor technologies. Color centers (CCs) in diamond moreover have long-lived spin ground states[8,10,11] and coherent optical transitions[3,12–15]; and group-IV-vacancy CCs in particular have stable optical transitions even in nanostructured diamond[12,16,17]. Their "symmetry-protected" resilience to electric field fluctuations has enabled a number of recent advances, including coherent optical transitions in nanophotonic waveguides[13,18–20] and cavities[21,22], memory-enhanced quantum communication[23], and a 2-qubit quantum network[24].

A central challenge now lies in scaling to large numbers of individually controllable color centers in a platform that simultaneously satisfies C1-C6[2]. Recent developments have shown individual optical addressing and capacitive frequency tuning for large numbers of waveguide-coupled color centers[20], but this architecture lacked methods for scalable, low-power-dissipation optical routing and coherent spin control. Conventional spin control methods using electromagnetic fields and microwave striplines suffer from high energy consumption and crosstalk[26,27]. Researchers recently showed promising alternatives that rely on the coupling of CC ground state levels to driven acoustic phonons in the host crystal[28–31]. However, methods reported to date rely on propagating acoustic modes excited in bulk samples via piezoelectric transducers[28], which present bottlenecks in component density and power dissipation under cryogenic conditions.

Here, we introduce an alternative device based on piezoelectrically driven nanoelectromechanical strain tuning of CCs in diamond nanophotonic waveguide arrays, or quantum microchiplets (QMCs). By driving localized acoustic modes in the diamond nanostructure, our "Strain-Transduction by Resonantly Actuated Integrated Nanoelectromechanical Systems" (STRAINEMS) device meets all essential criteria: (C2) strain-based frequency tuning of CC optical transitions; (C3) capability for coherent ground state control by exciting high-frequency acoustic modes; (C5) low device footprint to ~100 µm$^2$; and (C6) low power consumption due to localized resonance.

As shown in Figure 1, the STRAINEMS module completes the essential components of an architecture that can now satisfy all criteria C1-C6. We achieve this by hybrid integration of diamond nanophotonic waveguides with a cryogenically compatible System on Chip (SoC) that combines the STRAINEMS device with scalable optical addressing (C1) as well as high-speed electro-optic modulation and optical waveguide routing (C4)[32,33]. Based on full-wafer runs in a CMOS-compatible, 200 mm Si process[33,34], the SoC platform satisfies the scalable manufacturability requirement (C5). We term the full device a Hybrid Quantum System on Chip (HQ-SoC). While we focus on diamond color centers in this work, our architecture is widely applicable to heterogeneous integration of other solid state emitters such as quantum dots and defects in layered materials, limited by their sensitivity to strain.

## II. Nanoelectromechanical tuning of integrated color centers
### A. STRAINEMS structure and operation

In our approach, piezoelectric nano-cantilevers mechanically couple to a heterogeneously integrated QMC hosting implanted tin vacancy defects (SnVs, Fig. 1a). On-chip silicon nitride (SiN) waveguides optically couple to the QMC via inverse tapering[20,27], providing a scalable interface between SnV fluorescence and an active PIC. We excite the SnVs through free space, perpendicular to the QMC. The NEMs actuators share the same layer-stack as photonic modulators already demonstrated in our platform[32,34], allowing seamless integration of tunable quantum memories with large-scale photonics in the HQ-SoC. Trenches defining the undercut region of the cantilever confine the mechanical displacement, limiting crosstalk between actuators. This enables a compact device footprint without sacrificing operational frequency range.

Under an applied voltage $V(t) = V_{DC} + V_{AC}\sin(\omega_d t)$, the cantilever deflects along $\hat{Z}$ (Fig. 1c) introducing primarily uni-axial strain in the QMC along $\hat{X}$, $\varepsilon(t) \approx \varepsilon_{XX}(t) = \varepsilon_{XX,DC} + \varepsilon_{XX,AC}\sin(\omega_d t)$ (supplemental material section 1). This leads to a strain Hamiltonian with static and time varying components in the form of $H_{strain} = H_{static} + H_{AC}(t)$, where $H_{AC}(t)$ describes strain varying at a timescale faster than the CC radiative lifetime. Static strain shifts the optical transition energies of SnVs in the QMC by an amount $\Delta_n$ (where $n$ refers to a particular transition), while dynamic strain leads to sideband transitions at multiples of $\omega_d$[46].

The anisotropic nature of $\varepsilon(t)$ breaks the orientational degeneracy of SnVs in the QMC (Fig. 1d), with axial SnVs ([111], [$\bar{1}11$] dipole axes, purple) and transverse SnVs ([$\bar{1}11$], [$1\bar{1}1$] dipole axes, green) experiencing a distinct strain tensor (supplemental material figure 2) and correspondingly different deformation of their orbital states in response to $\varepsilon(t)$. Axial SnVs experience primarily a common mode shift in their optical transition energies due to strain along their dipole axis, while transverse SnVs experience relative shifts and state mixing due to off-axis strain[16,35,36]. The magnitude of $\Delta_n$ depends on the corresponding strain susceptibility parameter.

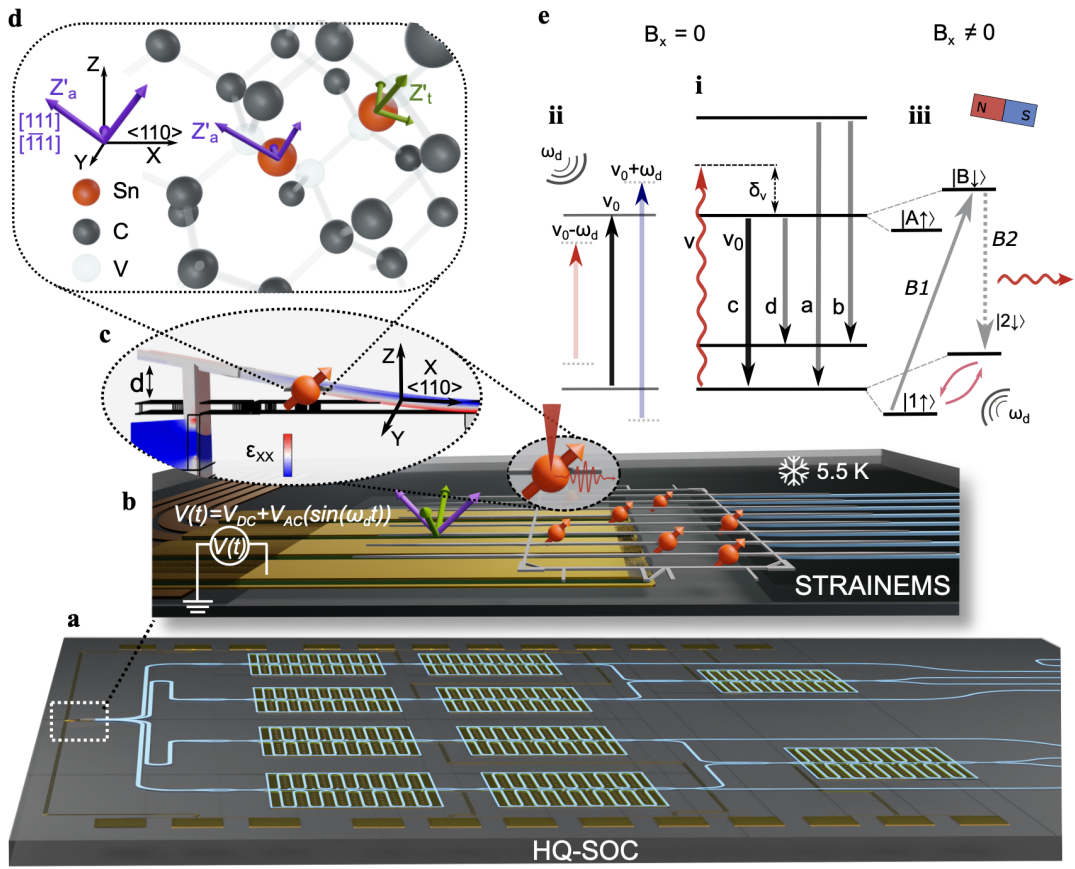

**Figure 1. a)** Hybrid quantum system on chip with components for quantum memory and control connected to active photonic modulators by optical waveguides (blue). **b)** Structure of the STRAINEMS component of the HQ-SoC integrating SnVs in a QMC with a piezoelectric cantilever and SiN waveguides for mechanical and optical coupling, respectively. **c)** Vertical deflection of the cantilever by an amount $d$ induces uniaxial strain in the attached QMC along the X axis (<110> crystal direction), $\varepsilon_{XX}$. **d)** SnV coordinate system (primed coordinates) relative to device coordinates and orientations within the diamond lattice. SnV dipole axes are aligned along the <111> crystal directions, with two distinct orientations relative to the uniaxial strain. Axial SnV dipole axes (purple, $Z'_a$) lie along [111] or [$\bar{1}11$], while transverse SnV dipole axes (green, $Z'_t$) along [$\bar{1}11$] or [$1\bar{1}1$]. **e) i)** SnV orbital states are split by the spin-orbit interaction, leading to four optical transitions (labeled 'a' through 'd') under zero external magnetic field. A laser field at frequency $\nu$ detuned from $\nu_0$ by an amount $\delta_\nu$ probes the $c$ transition. **ii)** Under acoustic driving by the cantilever when $\omega_d > \Gamma_{opt}$, red and blue detuned sidebands are visible in the PLE spectrum surrounding the central transition frequency $\nu_0$ at integer multiples of the drive frequency $\omega_d$ due to coupling with acoustic phonons. **iii)** An external magnetic field splits the spin-degenerate orbital states, allowing acoustic driving of the spin transition at $\frac{\omega_d}{2\pi}$ with optical readout using the spin-flipping B1 transition.

We probe the strain-dependence of SnV optical transitions using near-resonant laser excitation at frequency $\nu$ while monitoring the red-shifted phonon side-band (PSB). Fig.1ei illustrates this photoluminescence excitation (PLE) spectroscopy for probing shift $\Delta_c$ of SnV transition c.

For $\omega_d < \Gamma_{opt}$ (where $\Gamma_{opt}$ is the measured optical linewidth) we measure $\Delta_c$ directly from the shift in $\nu_0$ as a function of strain. In the resolved-sideband regime where $\omega_d > \Gamma_{opt}$, $\varepsilon(t)$ oscillates faster than the SnV radiative lifetime and we can observe $\varepsilon(t)$ as red and blue detuned sidebands in the PLE spectrum (Fig 1e ii). To investigate spin-phonon coupling of integrated SnVs, we place a permanent magnet near the sample to lift the spin degeneracy of the orbital ground states (Figure 1eiii). We observe PSB fluorescence under optical excitation of the spin-flipping B1 pathway, and apply an acoustic tone using the cantilever to drive transitions between the Zeeman split spin ground states.

## B. Frequency tuning of integrated tin vacancies

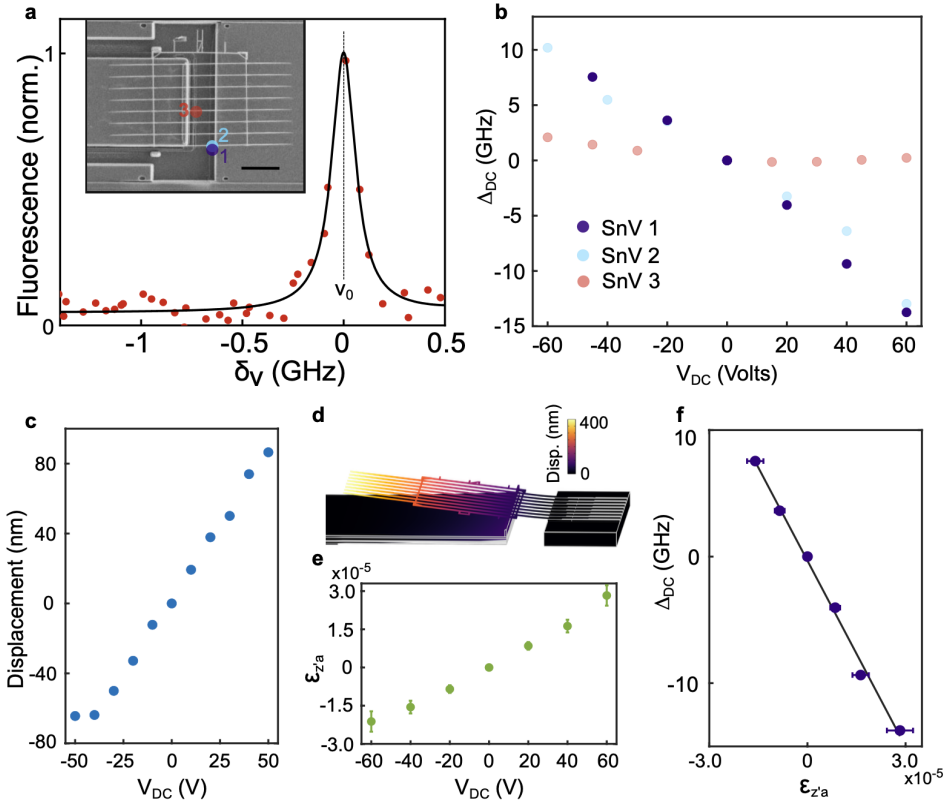

**Figure 2. a)** Photoluminescence excitation data for an SnV at the location marked by the red dot in the inset SEM image (SnV 3), taken at $\varepsilon(t) = 0$. **Inset:** Location of SnVs 1-3 within the QMC. **b)** $\Delta_{DC}$ vs $V_{DC}$ for SnV 1-3, with $V_{AC} = 0$. Scale bar is 10 $\mu m$. **c)** Displacement measured at the front edge of the cantilever vs $V_{DC}$. **d)** Calibrated finite-element model showing displacement consistent with measured values in panel c). **e)** $\varepsilon_{z'}$ extracted from finite-element simulations at the location of SnV1 at a depth of 75±5 nm (based on SRIM calculations) in the nanobeam cross-section, in the coordinate frame of an axial SnV. The uncertainty is due to implantation straggle estimated from SRIM calculations. **f)** $\Delta_{DC}$ vs $\varepsilon_{z'}$ for SnV1, fit using linear regression to extract $t_{\parallel,u} - t_{\parallel,g}$ of -0.490 ± 0.075 PHz/strain. The error bars in 2e and 2f are due to uncertainty in the SnV depth within the diamond nanobeam due to implantation straggle.

We first characterize $\Delta_c$ under static strain (referred to as $\Delta_{DC}$) with $V_{AC} = 0$. Figure 2a shows a PLE spectrum for an SnV at the location marked by the red dot in Figure 2ai when $\varepsilon(t) = 0$. We observe a linewidth of 120±18 MHz for this SnV extracted from a Lorentzian fit to the data (black curve), consistent with other SnVs in this device after integration and post-processing. When $V_{DC} \neq 0$, $\Delta_{DC}$ increases or decreases linearly depending on the sign of ε (Fig 2b, $\nu_0$ is obtained from Lorentzian fits to PLE data). Results from an SnV at a different location within the same QMC (SnV 3) and from the same location in a second STRAINEMS module with a different QMC (SnV 2) are also shown in Figure 2b. We consistently measure over 20 GHz frequency tuning for SnVs located in the region of the QMC under high DC strain (Supp. Fig 2), while the hold power dissipated on-chip for our device remains below 1 nW even at 60 V (Supplemental material section 4) satisfying requirement C6.

We next calibrate our finite-element model to accurately determine the strain in the diamond nanobeam and extract values for strain susceptibility $t_{\parallel,u} - t_{\parallel,g}$ for the SnV (we conjecture SnVs 1 and 2 are oriented axially based on the direction of their frequency shift vs strain). We adjust the adhesion between the QMC and the cantilever in our model until the simulated cantilever displacement (Fig. 2d) matches profilometer measurements of vertical deflection as a function of $V_{DC}$ (Fig. 2c). After extracting $\varepsilon_{Z'a}$ in the coordinate frame of an axially oriented SnV (Figure 2c), we fit $\Delta$ vs $\varepsilon_{Z'a}$ (Figure 2d) and obtain an estimate of -0.46±0.051 PHz/strain for $t_{\parallel,u} - t_{\parallel,g}$. We note that for an axial SnV $(\varepsilon_{x'x'} + \varepsilon_{y'y'}) \ll \varepsilon_{z'z'}$ (Fig 1c), and ignore the contribution from $t_\perp$ similar to the procedure used for SiVs and GeVs[35,36]. We find good agreement between the fits for SnVs located in high strain areas of the device (SnV1 and SnV2), yielding values of -0.490±0.075 and -0.436±0.071 PHz/strain respectively for $t_{\parallel,u} - t_{\parallel,g}$. While our estimated susceptibility values are slightly lower than those reported for the SiV and GeV (-1.7 PHz/strain), our estimate could be subject to additional errors arising from deviation of our finite-element model from the real device and variations in the depth of SnVs within the nanobeam cross-section.

# III. Resonant nanoelectromechanical control of color centers
## A. Frequency response and resonant enhancement

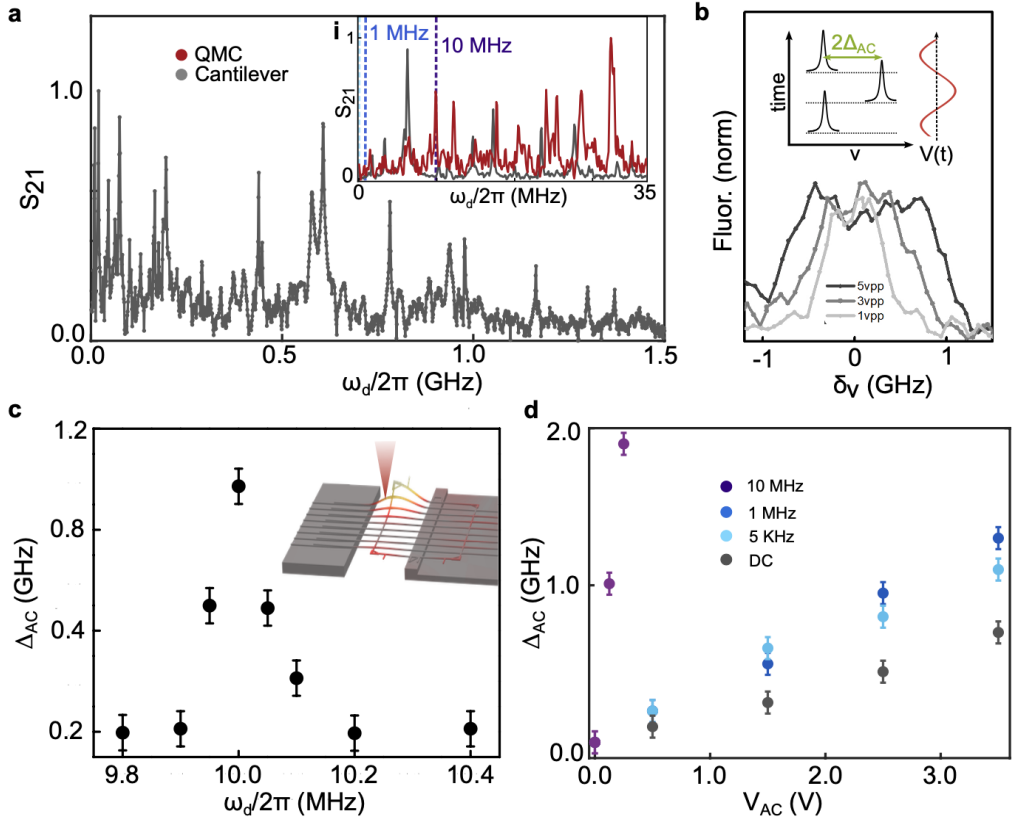

**Figure 3. a)** Frequency response of the device measured on the cantilever (gray curve) and QMC (red curve). **i)** Higher resolution frequency sweep of frequencies below 35 MHz. **b)** phonon sideband fluorescence as a function of $\delta_v$ for $V_{AC}$ of 0.5, 1.5, and 2.5 volts showing $\Delta_{ac}$ with $\omega_d/2\pi = 1\ MHz$. **c)** $\Delta_{ac}$ measured at $V_{AC} = 0.125\ V$ at $\omega_d/2\pi$ values from 9.8 to 10.4 MHz, showing the effect of a mechanical resonance at 10 MHz on $\Delta_{ac}$. inset: finite-element simulation of the mechanical mode responsible for the resonance at 10 MHz. **d)** $\Delta_{ac}$ measured for values both on resonance (10 MHz), and off resonance (DC, 5 kHz, 1 MHz).

We now apply a voltage $V_{AC} sin(\omega_d t)$ to the cantilever, and characterize the frequency-dependent behavior of the device. Confocal displacement measurements of the cantilever (Fig. 3a, 3a inset, gray curves) and coupled QMC (Fig. 3a inset, red curve) as a function of excitation frequency $\frac{\omega_d}{2\pi}$ reveal mechanical resonances extending to GHz frequencies (See supplemental section 4 for details of measurement setup). For $\frac{\omega_d}{2\pi} < \Gamma_{opt}$ we observe $\varepsilon(t)$ under AC driving as a broadened resonance in the PLE spectrum of SnVs, with a width equivalent to $2\Delta_{AC}$, where $\Delta_{AC}$ is the energy shift of the SnV transition at $V_{AC}$ (Fig. 3b).

The broadened linewidth provides a tool to explore the enhanced strain response of SnVs when the cantilever is driven at its mechanical resonances. Figure 3c plots $\Delta_{AC}$ measured with $V_{AC} = 0.125\ V$ for SnV 2, as we increase $\frac{\omega_d}{2\pi}$ through the mechanical resonance at 10 MHz in

Figure 3a. Voltage dependence reveals $\Delta_{AC}$ ~ 0.1 GHz for $\frac{\omega_d}{2\pi}$ values far from mechanical resonances (1 MHz, 5 kHz, and DC) vs 1.9 GHz at 0.25 $V_{AC}$ under resonant driving at 10 MHz, an almost 20-fold increase (Figure 3d). This amplified resonant response allows fast frequency tuning of integrated quantum memories with ultra low on-chip power dissipation. For the conditions in Figure 3, we estimate less than $4\times10^{-18}$ J switching energy (defined as the energy required to switch the device from $-V_{AC}$ to $+V_{AC}$) at 0.25 $V_{AC}$, or 0.4 nW at 10 MHz with a measured device capacitance of ~ 1 pF (Supplemental material section 3). Using the extracted susceptibility coefficient from DC measurements, we determine a switching energy of < 5 pJ/strain.

## B. Acoustic control of integrated tin vacancy centers

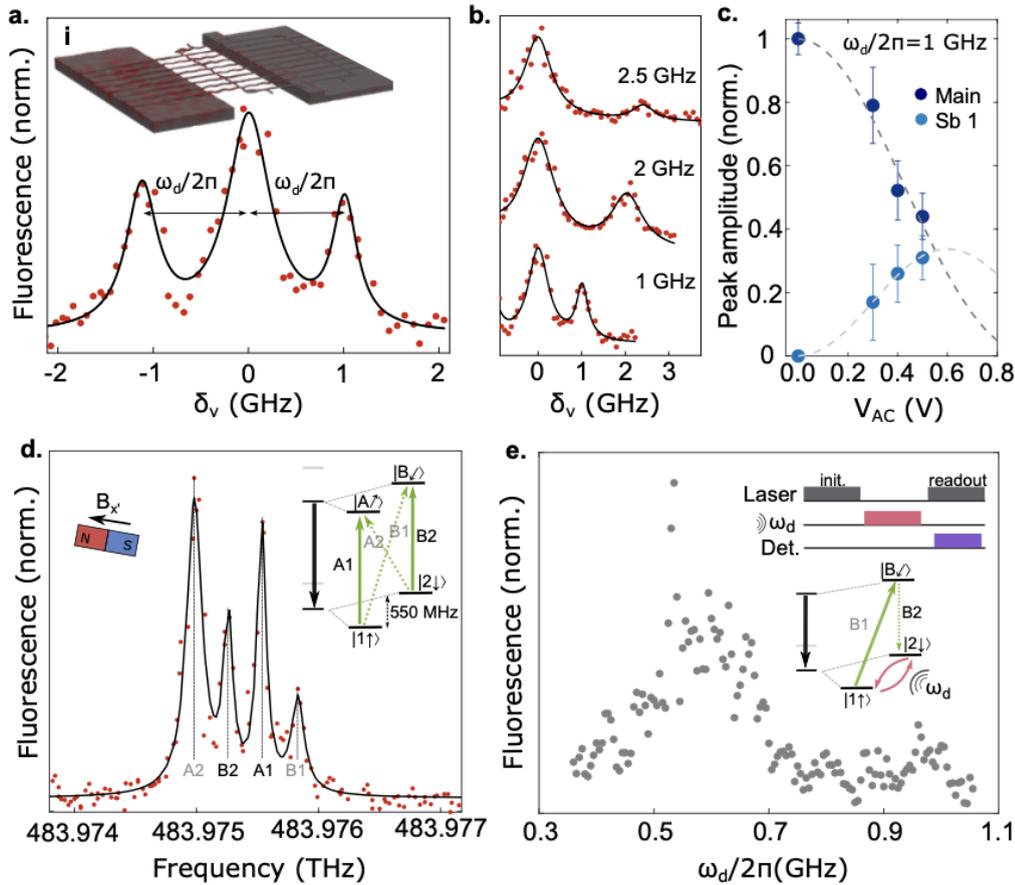

**Figure 4. a)** Phonon sideband fluorescence as a function of $\delta_v$ measured for $\omega_d/2\pi = 1\ GHz$, in the resolved sideband regime. Inset: FEM simulations of the mechanical deformation at 1 GHz. **b)** Fluorescence as a function of $\delta_v$ for $\omega_d/2\pi$ of 1, 2, and 2.5 GHz. **c)** Carrier peak and first sideband amplitudes as a function of $V_{AC}$, with error bars denoting 95% confidence intervals, derived from fits to the data. Dashed lines show a fit to Bessel functions of the first kind. **d)** Fluorescence as a function of ν in an external magnetic field, showing four distinct transitions due to Zeeman splitting of the spin states in the external magnetic field. **e)** Phonon sideband fluorescence as a function of $\omega_d/2\pi$ measured by exciting the spin flipping transition B1. **inset:** pulse sequence used for acoustic spin resonance measurements in main figure. A laser pulse resonant with B1 pumps the spin into |2>, followed by an acoustic pulse of frequency $\omega_d$. A readout pulse resonant with B1 detects spin population in state |1>.

The large operating frequency range of our device allows engineered coupling between acoustic vibrations and SnVs in the diamond nanostructure. We first investigate coupling between SnV orbital states and quantized strain in the vibrating nanobeam in the resolved-sideband regime, where $\frac{\omega_d}{2\pi} > \Gamma_{opt}$ and the response of the SnV to $\varepsilon(t)$ is described by $H_{AC}(t)$. The rapidly oscillating $\varepsilon(t)$ leads to coupling with virtual states visible as red- and blue-detuned sidebands in the PLE spectrum (Figure 4a, Snv 2) at integer multiples of the drive frequency $\frac{\omega_d}{2\pi}$. Figure 4ai shows a finite-element simulation of the mechanical mode responsible for the sideband occurring at 1 GHz (Fig. 4a, $V_{AC}$ = 0.5 V), arising from coupling with 1 GHz phonons. With $V_{AC}$ maintained at 0.5 V, we observe sidebands up to $\frac{\omega_d}{2\pi}$ = 2.5 GHz (Figure 4b), reflecting the large operational bandwidth of our device. The relative amplitudes of the sidebands and main peak fit to Bessel functions of the first kind (Figure 4c), with the population of the $k^{th}$ sideband given by[37]

$$P_k \approx |J_k(\frac{g_{orb}}{\omega_d} <n>)|^2 \qquad 1)$$

where $<n>$ is the phonon occupation number of the mechanical mode driven at $V_{AC}$, and $g_{orb}$ is the single-phonon coupling rate for the SnV orbital states, arising from strain due to zero-point fluctuations in the diamond nanobeam. Finite-element simulations of $g_{orb}$ yield ~ 2 kHz at the location of SnV 1(Supplemental section 5, Figure 8) with a maximum of ~ 8 kHz for an axially oriented SnV and ~10 kHz for transversely oriented SnVs. Using equation 1, we extract $<n>$ as a function of $V_{AC}$ (supplemental figure 10) and estimate $<n> \sim 10^5$ for $V_{AC} < 0.5\ V$. We note that under these conditions, the calculated on-chip dissipated power remains below 0.5 uW for frequencies exceeding the 2.5 GHz bandwidth of our device (supplemental material figure 5).

In order to investigate coupling between the SnV electron spin and phonons generated by the driven cantilever, we split the spin-degenerate orbital ground states of the SnV using a permanent magnet. We orient the magnetic field perpendicular to the SnV dipole axis for maximum spin-orbit mixing, and adjust the proximity of the magnet until the spin transition frequency extracted from the PLE spectrum (Figure 4d) roughly aligns with the ~600 MHz mechanical resonance observed in our devices (Supplemental section 5). We initialize the SnV spin to $|2\downarrow>$ by optically pumping the spin-flipping B1 transition, and probe the spin population in the $|1\uparrow>$ state following the application of a 150 ns acoustic pulse from the cantilever. Figure 4e shows the phonon sideband fluorescence under an optical readout pulse resonant with the B1 transition as the acoustic frequency is swept from 350 MHz to 1.05 GHz, following the pulse sequence shown in Fig. 4ei. When $\frac{\omega_d}{2\pi} \sim 550\ MHz$, at the frequency separation of the spin ground states (supplemental figure 10), we observe an increase in counts due to transfer of spin population from $|2\downarrow>$ back to $|1\uparrow>$ by the resonant acoustic pulse, suggesting successful manipulation of the SnV electron spin.

The efficiency of acoustic spin control in our device is limited by the coupling rate between the SnV spin and phonons generated by the driven cantilever. For SnVs the effective spin-phonon coupling rate, $g_{sm}$, depends on the degree of spin-orbit mixing as well as the pre-strain present in the sample. Using a model[12] taking into account magnetic field, pre-strain, and Jahn-Teller effects, we fit the peak locations in the PLE spectra shown in Figure 4d (Supplemental material figure 10), and estimate a transverse magnetic field of 0.022∓0.005T at the sample, with a ground state pre-strain of 865 GHz. Under these conditions, we narrow our estimate of $g_{sm}$ to $\sim 512\ Hz$.

# IV. HQ-SoC Operation

Application of our STRAINEMS module in future quantum networks and quantum processors motivates systems capable of both controlling integrated CCs at the nanoscale and routing their emitted photons, necessary for performing optically mediated entangling gates between many integrated spins. The HQ-SoC architecture integrates the STRAINEMS component with photonic modulators based on piezoelectric actuation, adding active photon routing capability. Figure 5a shows an optical microscope image of an example HQ-SoC comprising an 8-channel STRAINEMS module connected to two 4x1 switches. Each switch multiplexes four adjacent channels of the QMC to a single output (input) port, and allows interference between photons emitted from color centers in arbitrary combinations of the four input channels. The switches are composed of cantilever phase shifter MZIs (cps-MZIs) previously demonstrated in this platform[32]. Double cps-MZIs allow high contrast switching between adjacent input ports, with hardware error correction enabling >40 dB extinction in previous demonstrations[38]. The final single cps-MZI routes photons between output ports and functions as a beamsplitter for autocorrelation or interference measurements.

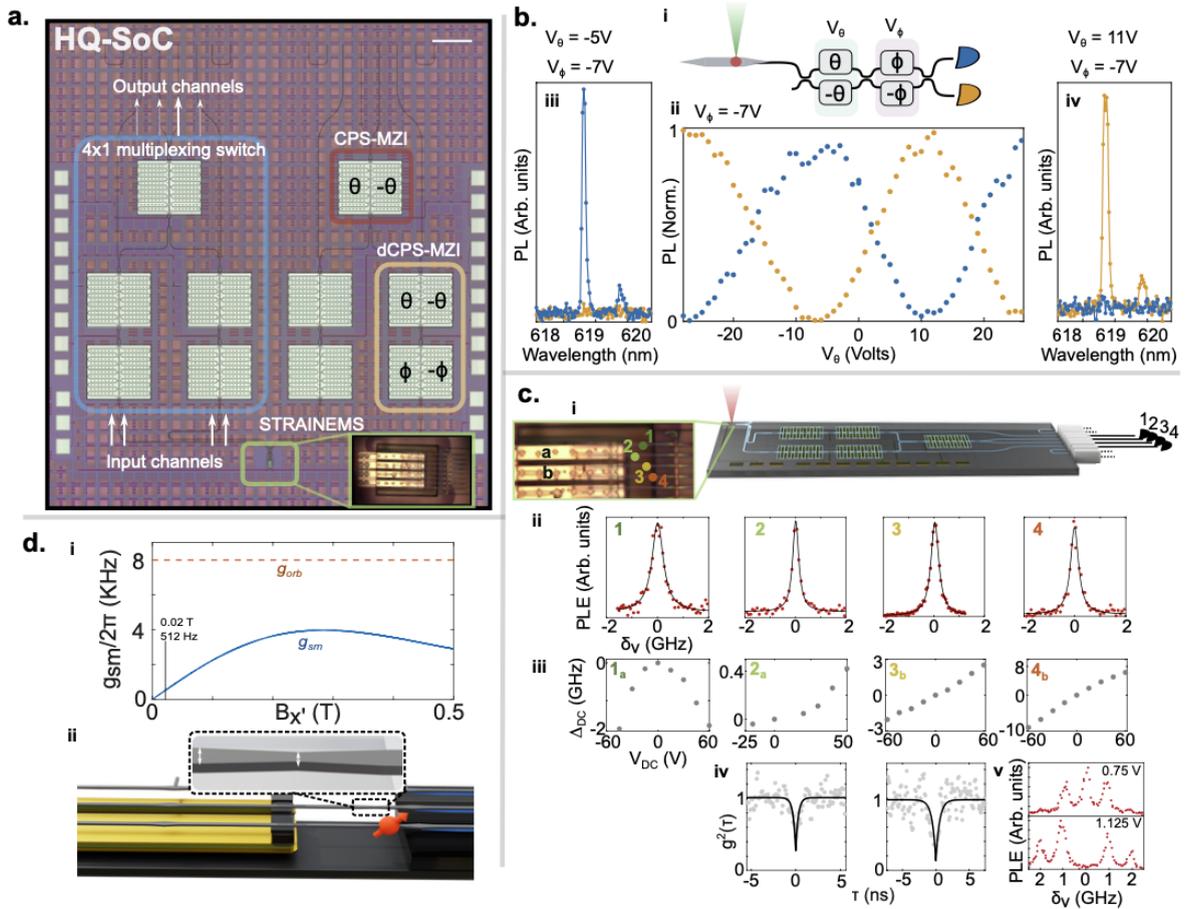

**Figure 5. a)** Optical microscope image of an HQ-SoC comprising an 8-channel STRAINEMS module connected to two 4x1 multiplexing switches. Double cantilever phase shifter MZI's (dCPS-MZI) allow high extinction switching between the four input channels. Single cantilever phase shifter MZIs (CPS-MZI) route light between output ports or function as a beamsplitter. **b)** The dCPS-MZI module allows high extinction switching of SnV fluorescence between the two output ports. **ii)** Normalized SnV fluorescence collected from the bar (blue) and cross (orange) ports of a dCPS-MZI with integrated QMC

as a function of $V_\theta$, while $V_\varphi$ is maintained at -7V. **iii, iv)** Fluorescence spectra measured at the bar and cross ports of the dCPS at $V_\theta$ = -7V, and 11V. **c)** Operation of a 4-channel switch within the HQ-SoC shown in panel a. **i)** Four channels of a fiber array collect light from the HQ-SoC and direct it out of the cryostat to single photon detectors labeled 1-4. Four outer channels are optionally used for optical alignment to the HQ-SoC waveguides. An optical microscope image shows a zoom-in to the STRAINEMS module with two actuators ("a" and "b") controlling CC's in channels 1-4 of the QMC. **ii)** PLE spectra for CC's in channels 1-4, collected at the output indicated in each plot. **iii)** $\Delta_{DC}$ for each emitter shown in ii) as $V_{DC}$ is applied to the indicated actuator. **iv)** Autocorrelation measurements for the CC's in channel 2 and 3 using the CPS-MZI to split the single photon output to detectors 2 and 3. **v)** PLE as a function of detuning with $\frac{\omega_d}{2\pi}$ for $V_{AC}$ = 0.75 V (top) and 1.2 V (bottom). **d) i.** Spin-phonon and orbital-phonon coupling rate as a function of transverse magnetic field, for an SnV with no pre-strain. **ii.** Engineering strain concentrating structures like constrictions in the nanobeams can help increase the zero-point strain, and raise the orbital coupling rate.

Figure 5b shows high contrast routing of SnV ZPL fluorescence under off-resonant excitation at 532 nm, using a dCPS-MZI connected to a single channel of an integrated QMC (supplemental figure 11). Control of both θ and φ in the dCPS-MZI (Fig 5b i) compensates for variations in non-tunable directional coupler splitting ratio[39] and allows >25 dB extinction of fluorescence from CCs (Figure 5b ii, Supplemental figure 11). We observe negligible SnV ZPL fluorescence from ports in the "off" state (Fig. 5b iii, iv) using control of both θ and φ to achieve maximum extinction.

Figure 5c demonstrates operation of the HQ-SoC shown in Figure 5a. Independent piezo-actuators within the STRAINEMS component each apply strain to two adjacent channels of the integrated QMC (Fig. 5a i). We consider CCs 1 through 4 marked in Figure 5ai, controlled by piezo-actuators 'a' and 'b'. Fluorescence from the CCs is routed through the 4x1 multiplexing switch into an edge-coupled multimode fiber array, and directed to single photon detectors (Fig. 5c i). Figure 5cii shows PLE spectra from the four CC's collected through output 2. All CCs show frequency tuning upon actuation of their cantilever (Figure 5c iii), with $\Delta_{DC}$ proportional to the simulated strain at their location within the QMC (see supplemental material section 2), up to a maximum of ~18 GHz for the high-strain region (CC 4).

We next use the cps-MZI at the output of the 4x1 switch as a 50:50 beam splitter to direct single photon emission to outputs 2 and 3. In this configuration, we perform an autocorrelation measurement for emitters 2 and 3, and observe an antibunching dip at $\tau = 0$ between signals collected from the two outputs (Fig. 5c iv). We conjecture that the background of 0.36±0.08 for emitter 2 and 0.16±0.05 for emitter 3 is due to background fluorescence from other emitters in the waveguide. Finally, we explore high frequency operation of the STRAINEMS module using actuator b and CC 4. Figure 5c v shows PLE spectra with $\frac{\omega_d}{2\pi} = 1\ GHz$, measured at increasing values of $V_{AC}$. We observe sidebands in the PLE spectra at integer multiples of 1 GHz up to 2 GHz, indicating coupling between CC orbital states and up to two phonons excited by the cantilever. This prototype device demonstrates the potential of the HQ-SoC architecture to combine active optical routing with photon frequency tuning and spin control, within a manufacturable integrated platform.

## V. Discussion and outlook

An important metric for acoustic manipulation of CC spins is the spin-phonon coupling rate, $g_{sm}$, since higher coupling rates enable faster and more efficient manipulation of the quantum memory spin degree of freedom. Many architectures in quantum networking and transduction rely on coherent coupling between spin states and single phonons[40–42], placing additional requirements on $g_{sm}$ and the mechanical behavior of the NEMS device. For SnVs in our STRAINEMS module, the effective

$g_{sm}$ results from an interplay between transverse magnetic field and static strain (pre-strain) present in the nanostructure. For samples with negligible pre-strain, $g_{sm}$ increases asymptotically towards $g_{orb}$ as a function of transverse magnetic field (Supplemental information figure 11). Conversely, the presence of pre-strain can counteract the effects of transverse magnetic field, quenching $g_{sm}$. The magnetic field dependence of $g_{sm}$ for the SnV measured in this work with ~860 GHz pre-strain reaches a maximum of ~4 kHz at around 0.3 T, and then decreases as a function of increasing magnetic field (Fig. 5di). In our measurements, the magnetic field strength was limited to 0.022 T due to using a permanent magnet located far from the sample, limiting $g_{sm}$ to ~500 Hz. Applying even modest transverse fields of 0.05-0.1 T, where the frequency of the SnV spin transition is still within the frequency range of our device (Supplemental figure 10a), could increase $g_{sm}$ to kHz frequencies without the need for complex mechanical resonator structures. For SnVs with minimal pre-strain, transverse magnetic fields beyond 0.5 T still yield spin transition frequencies within the bandwidth of our device (Supplemental figure 10b), allowing us to reach effective spin-phonon coupling rates near the orbital limit. Taking into consideration the phonon populations <n> estimated in the resolved sideband limit (Fig. 4c), bulk Rabi frequencies of over 100 MHz could be achieved under these conditions without the need for modification to the STRAINEMS design. Cooling our device to temperatures below 5.5 K (limited in this work by our cryostat) would allow coherent control of integrated SnV spins in a temperature regime with longer SnV spin coherence time, enabling us to characterize the Rabi frequencies and spin coherence times achievable in our device.

Beyond increasing $g_{sm}$ by modifying experimental conditions, we can further engineer the device to increase $g_{orb}$, raising the limiting constraint on the spin-phonon coupling rate. Increasing strain experienced by color centers due to zero-point motion of the structure can increase the orbital coupling rate and the efficiency of our device. Small modifications to the diamond nanobeam structure, for example engineering constrictions in the nanobeam width (Figure 5d ii), can confine the mechanical mode and enhance zero-point strain for color centers located in the concentrators. Similar structures have been shown to concentrate strain in optomechanical cavities[43], leading to MHz spin-phonon coupling rates.

## VI. Conclusion

In this work we demonstrate a hybrid system-on-chip architecture for strain control of color centers within an active PIC platform engineered for scalable optical addressing, modulation, and single-photon readout. The large bandwidth of the STRAINEMS component, spanning from DC to GHz frequencies, allows static strain tuning as well as AC driving and acoustic spin control. We measure frequency tuning near 25 GHz for color centers located in regions of high static strain, and measure a bandwidth of over 2 GHz while maintaining on-chip power dissipation below 1 $\mu$W. Resonant mechanical driving enables large estimated single phonon coupling rates near $10^4$ Hz for color center orbital states, and spin-phonon coupling rates approaching the orbital value under large transverse magnetic fields and minimal pre-strain (we estimate a value of $g_{sm}$~500 Hz at 0.022 T in our measurements). We demonstrate a prototype HQ-SoC device combining the STRAINEMS component with 4x1 multiplexing switches comprising piezoelectric photonic modulators, capable of routing single-photon emission from integrated color centers with high extinction. While we focused on color centers within a diamond host matrix in this work, our method is in principle widely applicable to other solid state quantum emitters and quantum materials, limited by their sensitivity to strain. Further refinement of our design to reduce pre-strain combined with operation in an optimized magnetic field could enable efficient acoustic spin control with kHz spin-phonon coupling rates, providing a scalable, integrated platform for quantum information processing and networking.

## Appendix A: Sample Preparation

We fabricate diamond quantum microchiplets from electronic grade diamond (Element Six) with implanted tin vacancy centers using a procedure described elsewhere[20]. Using a home built microscope setup equipped with a tungsten probe-tip (Ted Pella), we pick and place QMCs into PICs fabricated in a 200-mm foundry process at Sandia National Laboratory. After ensuring contact between the QMC and the STRAINEMS component we deposit a ~10 nm layer of $Al_2O_3$ using ALD, to glue the QMC to the PIC and promote mechanical coupling between the cantilever and the QMC. We have not observed any degradation of the optical properties of the color centres after the ALD step. We mount and wirebond the PIC with integrated QMC to a copper sample mount equipped with a PCB for applying electrical signals.

## Appendix B: Optomechanical measurements of SnV centers

We place the mounted PIC into a closed-cycle Montana cryostat with fiber feedthroughs for collecting fluorescence and electrical feedthroughs for applying DC and RF signals. Optical excitation and imaging is achieved using a home-built confocal microscope setup with excitation in free-space, perpendicular to the PIC. Fluorescence from color centers in the QMC is collected into an edge-coupled telecom lensed fiber and directed to single photon detectors (Perkin Elmer). For PLE measurements we use a tunable laser (M2 EMM) to resonantly excite the color centers alternating with an off-resonant 532 signal to reset the charge state, while collecting the phonon sideband using a long-pass filter during the resonant optical excitation. We apply DC electrical signals using a BK Precision power supply and use a Rigol 3 GHz signal generator to apply AC signals. We place a Nd permanent magent above the cryostat window near the sample to apply a magnetic field to the color centers, allowing us to tune the orientation and strength of the field by moving the magnet. This limits the strength of our magnetic field to 0.025 T however. In order to synchronize acoustic and optical pulses we use a Spincore PulseBlaster.

## Appendix C: HQ-SoC Measurements

For multi-channel collection from the HQ-SoC we use a multi-mode high numerical aperture 4-channel fiber array edge coupled to the PIC, while keeping the same confocal optical setup for excitation and imaging described in the previous section. We perform autocorrelation and antibunching measurements using an ID Quantique time tagger to histogram coincidence counts between two single photon detectors (Perkin Elmer), positioned at channels 2 and 3 of the fiber array. We use the same electrical setup to apply signals to the cantilevers in the STRAINEMS module as described previously, and use the BK precision power supply to apply signals to the CPS-MZIs in the HQ-SoC.

**Author Contributions**

GC built the cryo-optical setup, performed the optomechanical measurements, and designed the PICs. HR carried out FEM simulations of orbital-phonon and spin-phonon coupling rates. MK assisted with cantilever displacement measurements. KC and LL fabricated the diamond microchiplets. MZ and MD assisted with on-chip power dissipation measurement and calculations. YHW performed frequency response measurements. AL and DD fabricated the photonic chips. MT provided the model of SnV optical transitions and co-supervised HR. DE provided theory analysis. ME, GG, and DE supervised the work. GC and DE wrote the manuscript with input from all authors.

**Acknowledgements**

The authors acknowledge Dr. D. Andrew Golter for insightful discussions regarding this research. Major funding for this work is provided by The MITRE Corporation for the Quantum Moonshot Program. H.R. acknowledges support from the NDSEG Fellowship and the NSF Center for Ultracold Atoms (CUA). K.C.C. acknowledges additional funding support by the National Science Foundation RAISE-TAQS (Grant No.1839155). L.L. acknowledges funding from NSF QISE-NET Award (DMR-1747426) and the ARO MURI W911NF2110325. M.E. performed this work, in part, with funding from the Center for Integrated Nanotechnologies, an Office of Science User Facility operated for the US Department of Energy Office of Science. D.E. acknowledges the National Science Foundation (NSF) Engineering Research Center for Quantum Networks (CQN), awarded under cooperative agreement number 1941583.